\documentclass[pra,twocolumn,showpacs,aps,floatfix,amsmath,amssymb,amsfonts]{revtex4-1}

\usepackage{graphicx}
\usepackage{color}

\begin{document}

\title{Analytical model of overlapping Feshbach resonances}
\author{Krzysztof Jachymski$^{1}$ and Paul S. Julienne$^{2}$}
\affiliation{$^1$Faculty of Physics, University of Warsaw, Ho{\.z}a 69,
00-681 Warsaw, Poland,\\
$^{2}$Joint Quantum Institute, NIST and
the University of Maryland, Gaithersburg, Maryland 20899-8423, USA}
\date{\today}

\begin{abstract}
Feshbach resonances in ultracold collisions often result from an interplay between many collision channels. Simple two-channel models can be introduced to capture the basic features, but cannot fully reproduce the situation when several resonances from different closed channels contribute to the scattering process. Using the formalism of multichannel quantum defect theory we develop an analytical model of overlapping Feshbach resonances. We find a general formula for the variation of the scattering length with magnetic field in the vicinity of an arbitrary number of resonances, characterized by simple parameters. Our formula is in excellent agreement with numerical coupled channels calculations for several cases of overlapping resonances in the collisions of two $^7$Li atoms or two Cs atoms.
\end{abstract}

\maketitle
\section{Introduction}
Magnetically tunable Feshbach resonances have become an essential tool for the control of atomic interactions in ultracold quantum gases~\cite{Chin2010}. This is because such resonances allow the continuous tuning of the two-body $s$-wave scattering length, a key parameter that controls phenomena such as Bose-Einstein condensation~\cite{Inouye1998,Cornish2000}, the crossover physics between Bose-Einstein condensate and Bardeen-Cooper-Schriefer paring in fermonic gases of mixed species~\cite{Bourdel2004,Zwierlein2004}, and the formation of weakly bound Feshbach molecules~\cite{Jochim2003a,Lang2008}.  Near the pole position of a resonance at magnetic field $B_{res}$, the scattering length $a(B)$ takes on the following simple form as the field $B$ is varied~\cite{Moerdijk1995,Timmermans1999}:
\begin{equation}
  a(B) = a_{bg} \left ( 1 - \frac{\Delta}{B-B_{res}} \right ) \,,
  \label{eq:aB}
\end{equation}
where $a_\mathrm{bg}$ is a constant ``background'' scattering length in the region of the pole, and $\Delta$ is the width of the resonance; $a(B)=0$ at the ``zero crossing'' where $B = B_{res} + \Delta$. While scattering lengths are difficult to measure accurately, except for the locations of extreme features such as poles or zero crossings, they can generally be calculated accurately for alkali metal atom species using coupled channels calculations based on models of the ground state potentials~\cite{Chin2010}. An accurate $a(B)$ function is essential to interpret experiments involving three-body physics, since a key component of the theory is to know the scattering length at the laboratory field $B$~\cite{Kraemer2006,Knoop2008,Zaccanti2009,Khaykovich2010,Ferlaino2011, Berninger2011,Berninger2013,Dyke2013}.  

While Eq.~(\ref{eq:aB}) is applicable to an isolated single resonance, there are many cases of multiple overlapping resonances, and a generalization to such cases is needed.  We derive here an analytic representation of $a(B)$ due to a set of $N$ overlapping resonances from $N$ different closed channels interacting with a single open channel based on the Mies version of multichannel quantum defect theory (MQDT)~\cite{Mies1984a,Mies1984b,Julienne1989,Mies2000,Julienne2006,Julienne2009b}:
\begin{equation}
  a(B)=a_{bg}\left(1-\sum_{i=1}^N{\frac{\Delta_i}{B-B_i-\delta B_i -\sum_{j\neq i}{\frac{B-B_i}{B-B_j}\delta B_j}}}\right),
  \label{eq:aBgeneral}
\end{equation}
where the sum in the denominator represents a shift due to the mutual interaction of the resonances with the open scattering channel. We will show the connection between this formula and an alternative product form that is based on a square well model of multiple resonances~\cite{Lange2009}. 

This paper will give the MQDT framework for deriving Eq.~(\ref{eq:aBgeneral}) and illustrate it using fits to coupled channel calculations involving overlapping resonances relevant to the Efimov physics of colliding $^7$Li  atoms or Cs atoms.  We will discuss how mutual interaction between the resonances affects their positions and local widths. In particular, we will show how a narrow resonance on the shoulder of a broad one can be viewed as a ``local'' resonance in $B$ with a modified background and width due to the presence of the other resonance.

This work is organized as follows. In Sec.~II we briefly review the formulation of MQDT for van der Waals interactions. Sec.~III is devoted to derivation and analysis of the formulas describing a single Feshbach resonance using two-channel MQDT theory. In Sec.~IV we extend the model to the case of an arbitrary number of closed channels and resulting overlapping resonances. Sec.~V contains comparison of our theory to numerical and experimental results as well as characterization of some known resonances in lithium and cesium. Conclusions are drawn in Sec.~VI.

\section{Quantum defect theory.}
The collision of two cold atoms, including internal spin degrees of freedom and the effect of short-range forces, can generally be described by a multichannel matrix Schr\"{o}dinger equation
\begin{equation}
\frac{\partial \mathbf{F(r)}}{\partial r^2}+\frac{2\mu}{\hbar^2}(E\mathbf{I}-\mathbf{W}(r))\mathbf{F}(r)=0 \,,
\end{equation}
where $r$, $E$, $\mu$ and $\mathbf{I}$ are respectively the interatomic distance, energy, reduced mass, and the unit matrix.
References~\cite{Kohler2006,Chin2010} discuss the various kinds of basis sets that can be used to set up to describe the $(N+1)\times (N+1)$ interaction matrix $\mathbf{W}$ for the collision of two ground S state alkali metal atoms.  The number of scattering channels depends on the number of internal spin states of each atom and the number of orbital angular momenta needed to represent the collision.  The matrix $\mathbf{W}$ asymptotically approaches a diagonal form
\begin{equation}
W_{ij}\stackrel{r\to\infty}{\longrightarrow}\left(E^\infty_i+\frac{\hbar^2\ell_i(\ell_i+1)}{2\mu r^2}-\frac{C_6}{r^6}\right)\delta_{ij} \,.
\end{equation}
Here $\ell_i$ is the orbital angular momentum quantum number, $C_6$ is the coefficient of the van der Waals potential, and $E^\infty_i$ is the threshold energy for the $i$th channel. Depending on collision energy $E$, some of the channels can be open ($E>E^\infty_i$), and some are closed ($E<E^\infty_i$).   Here we assume a single $s$-wave open channel ($\ell=0$).

We start our analysis by a brief review of the MQDT formalism of Mies~\cite{Mies1984a,Mies1984b,Mies2000}, which offers some different physical insights than alternative MQDT formalisms~\cite{Burke1998,Gao2008,Gao2011}. Because of the long-range van der Waals potential,  we will like Gao~\cite{Gao2008,Gao2011} use $R_6=(2\mu C_6/\hbar^2)^{1/4}$ as the unit of length and $E_6=\hbar^2/2\mu R_6 ^2$ as the unit of energy.   In MQDT one replaces $\mathbf{W}$ by a set of reference potentials which reproduce the asymptotic form of $\mathbf{W}$ and parameterizes the solution of the new diagonal reference problem at short distances using functions with WKB-like normalization $\hat{f}$, $\hat{g}$. These functions are connected with the long-distance scattering solutions $f$, $g$ (for open channels) or exponentially decaying solutions $\phi$ (for closed channels) using the MQDT functions $C(E)$, $\tan\lambda(E)$ and $\nu(E)$~\cite{Julienne1989,Mies2000,Julienne2009b,NJP2011}
\begin{align}
\begin{array}{lll}
f_i(r)    & = & C_i^{-1}(E)\hat{f}_i(r)\\
g_i(r)    & = & C_i(E)(\hat{g}_i(r)+\tan\lambda_i\hat{f}_i(r))\\
\phi_i(r) & = & \mathcal{N}(E)(\cos\nu_i(E)\hat{f}_i(r)-\sin\nu_i(E)\hat{g}_i(r)),
\end{array}
\end{align}
where $\mathcal{N}(E)$ ensures unit normalization of the bound state function; see Ref.~\cite{Julienne2009b} for a discussion of the normalization of the MQDT functions. Eigenvalues $E_{ni}$ of the reference closed channel $i$ occur where $\tan\nu_i(E_{ni})=0$.   The general solution of the coupled channels problem can be written as
\begin{equation}
\mathbf{F}(r)=(\mathbf{f}^0(r)+\mathbf{Y}\mathbf{g}^0(r))\mathbf{A},
\end{equation}
where $\mathbf{f}^0(r)$ and $\mathbf{g}^0(r)$ are diagonal matrices of the reference solutions, $\mathbf{Y}$ is the quantum defect matrix which contains information about the short-range couplings, and $\mathbf{A}$ gives the amplitudes. The short-range processes which determine $\bf{Y}$ are assumed to be present only at length scales $R_0\ll R_6$. This brings an important simplification to the problem, as the energy scales associated with the short range are much bigger than $E_6$, which is of the order of milikelvins. As a result, for ultracold collisions the $\mathbf{Y}$~matrix can be regarded as energy-independent.   Moreover, assuming that the potential varies from its long-range form only at short distances makes it possible to use the analytic theory of van der Waals interactions for the MQDT functions~\cite{Gao1998a,Gao2000}.

The observable properties of the system are given in terms of the open channel block of the scattering matrix~$\mathbf{S}$. Within the framework of MQDT, it can be obtained from the quantum defect matrix $\mathbf{Y}$ and the open channel quantum defect functions, as follows~\cite{Mies1984a}:
\begin{equation}
\label{So}
\mathbf{S}_{oo}=e^{i\mathbf{\xi}_{oo}}(1+i\mathbf{R}_{oo})(1-i\mathbf{R}_{oo})^{-1}e^{i\mathbf{\xi}_{oo}},
\end{equation}
where $i\mathbf{\xi_{oo}}$ is a diagonal matrix with elements $\xi_i\delta_{ij}$ giving the phase shifts for the reference potentials, and
\begin{equation}
\label{Ro}
\mathbf{R}_{oo}=\mathbf{C}^{-1}(E)(\bar{\mathbf{Y}}_{oo}^{-1}-\tan\mathbf{\lambda}(E)_{oo})^{-1}\mathbf{C}^{-1}(E) \,.
\end{equation}
The matrices of the reference channel quantum defect functions $\tan\lambda(E)$, $C(E)$ and $\tan\nu(E)$ are diagonal.
The renormalized open channel $\bar{\mathbf{Y}}_{oo}$ matrix is
\begin{equation}
\label{Yren}
\bar{\mathbf{Y}}_{oo}=\mathbf{Y}_{oo}-\mathbf{Y}_{oc}(\tan\mathbf{\nu}(E)_{cc}+\mathbf{Y}_{cc})^{-1}\mathbf{Y}_{co} \,,,
\end{equation}
where the indices $o$ and $c$ respectively denote open and closed channels.  Here we assume a single open channel.

\section{MQDT description of a two-channel resonance.}
We will now provide the MQDT description of a single magnetically tunable Feshbach resonance in the simplest case when only one open and one closed channel are present~\cite{Julienne2006}.  This is general, since Mies {\it et al.}~\cite{Mies2000a} showed how a problem with a single isolated resonance due to multiple closed channels can be reduced to a problem with a single effective closed channel; see also Ref.~\cite{Nygaard2006}.   By choosing the reference potentials to reproduce the true scattering lengths of the uncoupled open and closed channels, the quantum defect matrix $\mathbf{Y}$ will contain only off-diagonal terms
\begin{equation}
\mathbf{Y}=\left(
\begin{array}{ll}
0&y\\
y&0
\end{array}
\right) \,.
\end{equation}
The dimensionless short range coupling parameter $y$ is assumed to be independent of $E$ and $B$.
Substituting this into equations \eqref{So}-\eqref{Yren}, we obtain the $\mathbf{S}_{oo}$~matrix (which in the case of a single open channel is just a complex number $S$) in a factored form with the two factors representing the background and resonant scattering parts:
\begin{equation}
\label{2chS}
S=e^{2i\xi}\left(1-\frac{2iy^2C^{-2}(E)}{\tan\nu(E)+y^2\tan\lambda(E)+iy^2C^{-2}(E)}\right)\,.
\end{equation}
The open reference channel phase shift $\xi \to -ka_{bg}$ as $k \to 0$, and for convenience we choose to write the background scattering length in our $R_6$ length units as $a_{bg}=r\bar{a}$, using the mean scattering length of the van der Waals potential, $\bar{a}=2\pi/\Gamma\left(\frac{1}{4}\right)^2 \approx 0.478$, introduced by Gribakin and Flambaum~\cite{Gribakin1993}.

The $C^{-2}(E) $ and $\tan\lambda(E) $ functions control the threshold behavior of resonance scattering, giving the respective amplitude and phase relations between the short- and long-range reference functions.  Discussion of the physical meaning of these MQDT functions in the cold collision context is given in Refs.~\cite{Julienne1989,Mies2000,Julienne2006,Julienne2009}.  When the collision energy $E=\hbar^2 k^2/(2\mu)$ in the open channel becomes large compared to $E_6$, then $C^{-2}(E) \to 1$ and $\tan\lambda(E) \to 0$.  The threshold behavior for $s$-waves as $E \to 0$ is~\cite{Mies2000}
\begin{equation}
\label{qdtfunctions}
 C^{-2}(E) \to k \bar{a} \left ( 1 + (1-r)^2 \right ), \,\,\,\,\,  \tan\lambda(E) \to 1-r \,.
 \end{equation}

The $\tan\nu(E)$ function in Eq.~(\ref{2chS}) vanishes at an eigenvalue $E=E_n$, where the quantum number $n$ labels the vibrational level. Near such an eigenvalue we can expand the energy-dependence as~\cite{Mies1984b}
\begin{equation}
\label{tannu}
\tan\nu(E) \approx \left (\frac{\partial \nu}{\partial E}\right)_{E=E_n} \left [ E-\delta\mu(B-B_1) \right ]
\end{equation}
where $E_n=\delta\mu(B-B_1)$ is the field-dependent position of the ``bare'' closed channel eigenvalue, $\delta\mu$ is the magnetic moment difference between the bare open and closed channel states (measured in $E_6$ per gauss units), and $B_1$ is the magnetic field at which the bare bound state crosses the open channel threshold. Note that $\pi (\partial E/\partial \nu)_{E_n} = \delta E_n$ is the mean vibrational spacing near $E=E_n$, and $\delta E_n/h$ is the corresponding vibrational frequency.  If $n$ does not represent the last bound state of the closed channel, a useful approximation is $\delta E_n \approx (E_{n+1} - E_{n-1})/2$.

It is straightforward to relate the expression in Eq.~(\ref{2chS}) to conventional resonant scattering by rewriting the resonant factor as
\begin{equation}
\label{resterm}
 1-\frac{i \hat{\Gamma}C^{-2}(E)}{E-\delta\mu(B-B_1)+\frac12\hat{\Gamma}\tan\lambda(E)+i\frac12\hat{\Gamma}C^{-2}(E)} \,,
\end{equation}
where
\begin{equation}
\label{Gammahat}
  \hat{\Gamma} = 2 y^2  \left (\frac{\partial E}{\partial \nu}\right)_{E=E_n}
\end{equation}
is a constant that represents a short-range decay width.  Note that the form in Eq.~(\ref{resterm}) gives the entire $B$-dependence in the $B-B_1$ term from the expansion of $\tan\nu(E)$, while the $C(E)^{-2}$ and $\tan\lambda(E)$  MQDT functions give the near-threshold variation with energy.  The two terms in the denominator proportional to $\hat{\Gamma}$ represent the energy-dependent shift and width due to the threshold resonance.  The numerator of the pole term in Eq.~(\ref{resterm}) represents the threshold decay width
 \begin{equation}
   \Gamma(E) = 2 \pi |\langle \phi_n | W_{oc}(r) | f_o \rangle |^2 = \hat{\Gamma} C^{-2}(E) 
 \end{equation}
where $W_{oc}(r)$ is the short-range coupling between the bare open and closed channels.  Thus, $\hat{\Gamma}$ represents the width when the open channel scattering wave function $f(r)$ is replaced by the $\hat{f}(r)$ wave function with short range WKB normalization.   Following Ref.~\cite{Mies1984b}, when $y^2 \ll 1$ the quantity $4y^2$ can be interpreted as the short-range probability that the bound state decays into the open channel during a single vibrational cycle. Multiplying by $C^{-2}(E)$ converts the short-range probability into the proper threshold probability.   Consequently, the bare bound state decay rate $\Gamma(E)/\hbar=(4y^2C^{-2}(E))(\delta E_n/h )$ can be interpreted as the decay probability per cycle times the closed channel vibrational frequency (number of cycles per second).   
 
Simple algebraic transformations of (\ref{2chS}) give the energy-dependent scattering length~\cite{Bolda2002,Krych2009,Idziaszek2010}, defined as $\frac{1}{ik}\frac{1-S}{1+S}$, in the form
\begin{equation}
\label{aen}
a(E,B)=\frac{C^{-2}y^2/k-(\tan\nu+y^2\tan\lambda)\tan\xi/k}{\tan\nu+y^2\tan\lambda+C^{-2}y^2\tan\xi} \,.
\end{equation}
By taking the $E\to 0$ limit and making use of relations~\eqref{tannu}-\eqref{Gammahat}, we obtain the standard scattering length
\begin{equation}
\label{ascat}
a(B)=a_{bg}-\frac{\frac{1}{2}\left.\hat{\Gamma}C^{-2}(E)/k\right|_{E\to 0}}{\delta\mu(B-B_1)-\frac{1}{2}\left.\hat{\Gamma}\tan\lambda(E)\right|_{E\to 0}} \,.
\end{equation}
We can now {\it define} a resonance width $\Delta$ by writing the  ``pole strength'' in the numerator in Eq.~(\ref{ascat}) as 
\begin{equation}
\label{s6}
 s_6 = a_\mathrm{bg} \Delta \delta \mu = \frac12 \hat{\Gamma} C^{-2}(E)/k
 \end{equation}
Note that the $\Delta$ so defined is the same as in Eq.~(22) of Chin {\it et al.}~\cite{Chin2010};  the dimensionless resonance strength parameter defined by Chin {\it et al.}~\cite{Chin2010} is $s_\mathrm{res} =\bar{a} s_6 \approx 0.478 s_6$.  The $s_\mathrm{res}$ (or $s_6$) parameter is very important for characterizing the properties of an isolated resonance and also may turn out to be of particular importance in the analysis of three-body recombination near the resonance~\cite{Wang2013}.

With the above definitions, and using the threshold properties in Eq.~(\ref{qdtfunctions}), 
the $s$-wave scattering length is given by the familiar formula~\cite{Moerdijk1995,Timmermans1999,Chin2010}
\begin{equation}
\label{asingle}
a(B)=a_{bg}\left(1-\frac{\Delta}{B-B_1-\delta B}\right),
\end{equation}
where the width and shift are proportional to $y^2$:
\begin{eqnarray}
\label{Delta}
  \Delta &=& \frac{\delta E_n}{\pi \delta \mu} y^2 \frac {1+(1-r)^2}{r} = \frac{s_6}{a_{bg} \delta \mu}\\
  \label{DeltaB}
  \delta B &=& \frac{\delta E_n}{\pi \delta \mu} y^2 (1-r) = \Delta \frac{r(1-r)}{1 + (1-r)^2}
\end{eqnarray}
The latter formula for the shift, previously given in Refs.~\cite{Kohler2006,Julienne2006,Chin2010} without derivation, shows that the shift can not be much larger in magnitude than $|\Delta|$, towards which it tends for large $|r|$.

\section{Many closed channels.}
 In many physical systems coupling to a single closed channel is not sufficient to describe the scattering and bound states. In fact, coupled-channels calculations often show many overlapping resonances due to couplings with several closed channels which have poles near one another as a function of $B$~\cite{Takekoshi2012,Berninger2013,Gross2011}. It is noteworthy that several resonances that have been used to study exotic three-body physics and the Efimov effect occur in regions with overlapping resonances~\cite{Berninger2011,Gross2011}.  When resonances appear near one another, the simple formula~\eqref{asingle} fails to describe the scattering length properly. However, it is straightforward to extend the results from the previous section by adding additional closed channels to the model. We thus start from the quantum defect matrix of the form
\begin{equation}
\mathbf{Y}=\left(
\begin{array}{llll}
0&y_1&y_2&\ldots\\
y_1&0&0&\ldots\\
y_2&0&0&\ldots\\
\ldots&\ldots&\ldots&\ldots
\end{array}
\right).
\end{equation}
It is in fact usually not necessary to include couplings between the closed channel states. We can instead assume that we use a basis in which the closed channels are already diagonalised~\cite{Mies1968}. Then by choosing the reference potentials to reproduce the scattering lengths we get rid of the diagonal terms in the quantum defect matrix as well.  As a result of a similar procedure as before and simple algebraic transformations, we obtain the energy-dependent scattering length
\begin{equation}
\label{aenmany}
ka(E)=\frac{C^{-2}\sum_i{\frac{y_i^2}{\tan\nu_i}}-\tan\xi\left(1+\tan\lambda\sum_i{\frac{y_i^2}{\tan\nu_i}}\right)}{C^{-2}\tan\xi\sum_i{\frac{y_i^2}{\tan\nu_i}}+1+\tan\lambda\sum_i{\frac{y_i^2}{\tan\nu_i}}}\, ,
\end{equation}
where the dependence on $B$ is contained in the expansion of each $\tan\nu_i(E)$ terms as in Eq.~(\ref{tannu}).
The standard scattering length as $E \to 0$ is
\begin{equation}
\label{ageneral}
a(B) = a_\mathrm{bg} - \sum_{i=1}^N P_i(B) \,.
\end{equation}
where the resonant pole terms,
\begin{equation}
\label{ageneralpole}
P_i(B)={\frac{\frac{1}{2}\frac{\hat{\Gamma}_i}{\delta\mu_i}C^{-2}(E)/k}{B-B_i-\frac{1}{2}\tan\lambda(E)\left(\frac{\hat{\Gamma}_i}{\delta\mu_i}-\sum_{j\neq i}{\frac{B-B_i}{B-B_j}\frac{\hat{\Gamma}_j}{\delta\mu_j}}\right)}} \,,  \nonumber
\end{equation}
imply the $E \to 0$ limit,  the widths $\hat{\Gamma}_i = 2y_i^2 (\delta E_{n\,i}/\pi)$ are defined as in Eq.~(\ref{Gammahat}), and $C(E)^{-2}$, $\tan\lambda(E)$, and $r$ are defined for the open channel with background scattering length $a_{bg}$.   By defining a resonance width $\Delta_i$ and shift $\delta B_i$  for each resonance as in Eqs.~(\ref{s6}) and (\ref{DeltaB}), we find
\begin{equation}
\label{ascg}
a(B)=a_{bg}\left(1-\sum_i{\frac{\Delta_i}{B-B_i-\delta B_i-\sum_{j\neq i}{\frac{B-B_i}{B-B_j}\delta B_j}}}\right).
\end{equation}
The mutual influence of the resonances on one another is thus contained in the terms $\frac{B-B_i}{B-B_j}\delta B_j$. We note that this influence is not due to direct coupling between the closed channels, but rather to indirect interaction via the open channel.

Several important comments are in order here. Firstly, from looking at the structure of Eq.~\eqref{ascg} one may suppose that there may be no simple local parameter similar to $s_{res}$ any more, since all the widths are needed to fully describe each resonance. However, it is possible to rewrite Eq.~\eqref{ascg} in the form of isolated terms with new parameters, so that it is possible to define a meaningful $s_{res}$ parameter for each resonance. One particularly interesting case is the interplay between a broad resonance and a very narrow one. This is a fairly common situation if a weak resonance of high partial wave character exists near a broad $s$-wave resonance. Assuming two resonances with $\left|\Delta_1\right|\ll\left|\Delta_2\right|$, Eq.~\eqref{ascg} can be simplified to
\begin{equation}
\label{narrow}
a(B)\approx a_{bg}\left(1-\frac{\Delta_2}{B-B^{res}_2}-\frac{\alpha \Delta_1}{B-B^{res}_1}\right),
\end{equation}
where $\alpha=\left((B_1-B_2)/(B_1-B_2-\delta B_2)\right)^2$ renormalizes the width of the narrow resonance and $B^{res}_i$ denotes the two pole positions. The new width can be either larger or smaller than $\Delta_1$, depending on the background scattering length and the relative position of resonances. One can treat the narrow resonance as an isolated one and describe it by the $s_{res}$ parameter using the renormalized width.  By rewriting Eq.~(\ref{narrow}) as
\begin{equation}
\label{alocal}
        a(B) = a'_{bg} \left ( 1 - \frac{\Delta'_1}{B - B_1^{res}} \right ) \,,
\end{equation}
the scattering length can be approximated as a ``local'' isolated narrow resonance near $B \approx B_1^{res}$ with
\begin{equation}
 a'_{bg} = a_{bg} \left (1-\frac{\Delta_2}{B_1^{res} -B_2^{res}} \right ) ,  \,\,\,\,     \Delta'_1 = \frac{\alpha \Delta_1}{1 - \frac{\Delta_2}{B_1^{res} - B_2^{res}}},
\end{equation}
where $s_6/\delta\mu_1 = a'_{bg}\Delta'_1 = a_{bg} \alpha \Delta_1$ is the same whether the ``local''  background $a'_{bg}$ or ``global'' background $a_{bg}$ is used.  

In general, it is also possible to algebraically transform formula~\eqref{ascg} into a product form which was previously derived using a simple model of coupled square wells~\cite{Lange2009} 
\begin{equation}
\label{achin}
a(B)=a_{bg}\prod_{i=1}^N{\left(1-\frac{\tilde{\Delta}_i}{B-B^{res}_i}\right)}.
\end{equation}
The transformation can be done by noticing that both~\eqref{ascg} and~\eqref{achin} may be represented in the form $a_{bg}\left(1-w_1(B)/w_2(B)\right)$, where $w_1$ and $w_2$ are polynomials of $(N-1)$th and $N$th order in $B$. By equating the coefficients in the polynomials one obtains a set of equations connecting both formulas, which can be solved numerically. The resonance positions $B^{res}_i$ are given by the zeros of the denominators in~\eqref{ascg}. However, this usually does not determine the $\tilde{\Delta}$ parameters uniquely. The most ``intuitive'' solution gives $\tilde{\Delta}_i$ as the distance between the pole and the nearest zero of the scattering length, but other solutions are also possible.  Consequently, there is no clear interpretation of the $\tilde{\Delta}_i$ parameters as resonance widths. Remarkably, each $\tilde{\Delta}_i$ and resonance position $B^{res}_i$ is a function of \textit{all} the bare widths $\Delta_i$, crossing positions $B_i$ and the background scattering length in Eq.~(\ref{ascg}). Thus, one can always use Eq.~\eqref{achin} to define a pole strength for a ``local'' pole $i$ as in Eq.~(\ref{alocal}) as a product of a ``local'' width $\tilde{\Delta}_i$ and a ``local'' background scattering length,
\begin{equation}
\tilde{a}_{bg,i} =a_{bg}\prod_{j\ne i}^N{\left(1-\frac{\tilde{\Delta}_j}{B^{res}_i-B^{res}_j}\right)} \,.
\end{equation}
Both $\tilde{a}_{bg,i}$ and $\tilde{\Delta}_i$ are functions of all the other pole terms (that is, all the $y_i$ parameters), and only their product remains well-determined when the fitting range eliminates distant poles that affect the local region.

Another important question is how to extract multiple resonance parameters by fitting numerical coupled channel calculations of $a(B)$. In some cases, such as $^7$Li in the $f=1,\,m_f=0$ spin channel, there are only two resonances~\cite{Gross2011}, and fitting the formula~\eqref{ascg} is relatively easy. In other cases, such as cesium in its $f=3,\,m_f=3$ spin channel, the number of resonances is very high~\cite{Berninger2013}, and the fitting becomes computationally costly.  
Excluding some resonances from the fitting will matter for the uniqueness of the fit, as Eq.~(\ref{ascg}) is nonseparable. The effect is larger if any omitted resonances are quite broad or if they lie close to the region of interest. A parameter,
\begin{equation}
\label{beta}
\beta_i=\frac{\tilde{a}_{bg,i} \tilde{\Delta}_i}{a_{bg}\Delta_i} \,,
\end{equation}
which can be numerically determined by separately fitting $a(B)$ to Eqs.~(\ref{ascg}) and (\ref{achin}),
can be introduced as a measure of the impact of other resonances on the $i$th one. In any case, the product $\tilde{a}_{bg,i} \tilde{\Delta_i}$ obtained from the fitting tends to be robust, even if the individual terms are not.  Consequently, a meaningful pole strength $\tilde{s}_6 = \tilde{a}_{bg,i}\tilde{\Delta}_i \delta \mu_i$ can be defined for each resonance.

Finally, it should be noted that our general formula in Eq.~(\ref{aenmany}) shows how to obtain the scattering properties at finite energies away from $E=0$.  Thus, it is more general than the expressions in Eqs.(\ref{ascg}) or (\ref{achin}) and can yield effective range or other, more rigorous, finite energy corrections in the presence of single or multiple resonances.

\section{Applications}

\begin{figure}
\centering
\includegraphics[width=\linewidth]{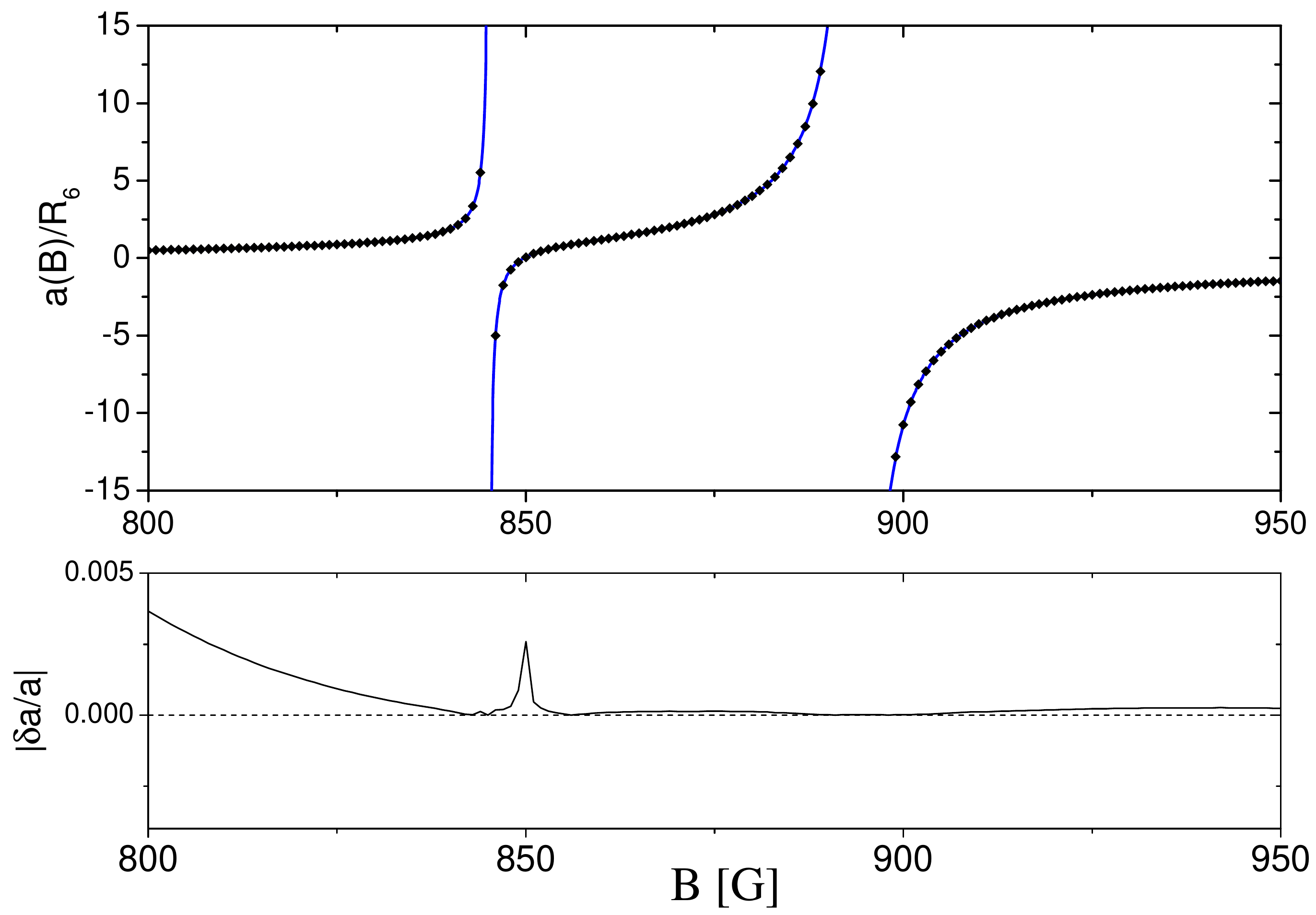}
\caption{\label{li}(color online) Top: two $s$-wave Feshbach resonances for the collision of two $f=1,\,m_f=0$ $^7$Li atoms. Coupled-channel calculations (black points) compared with the fitted formula~\eqref{ascg} (blue line). Bottom: the relative deviation of the fit from the numerical calculations.}
\end{figure}

\begin{figure*}
\centering
\includegraphics[width=0.31\linewidth]{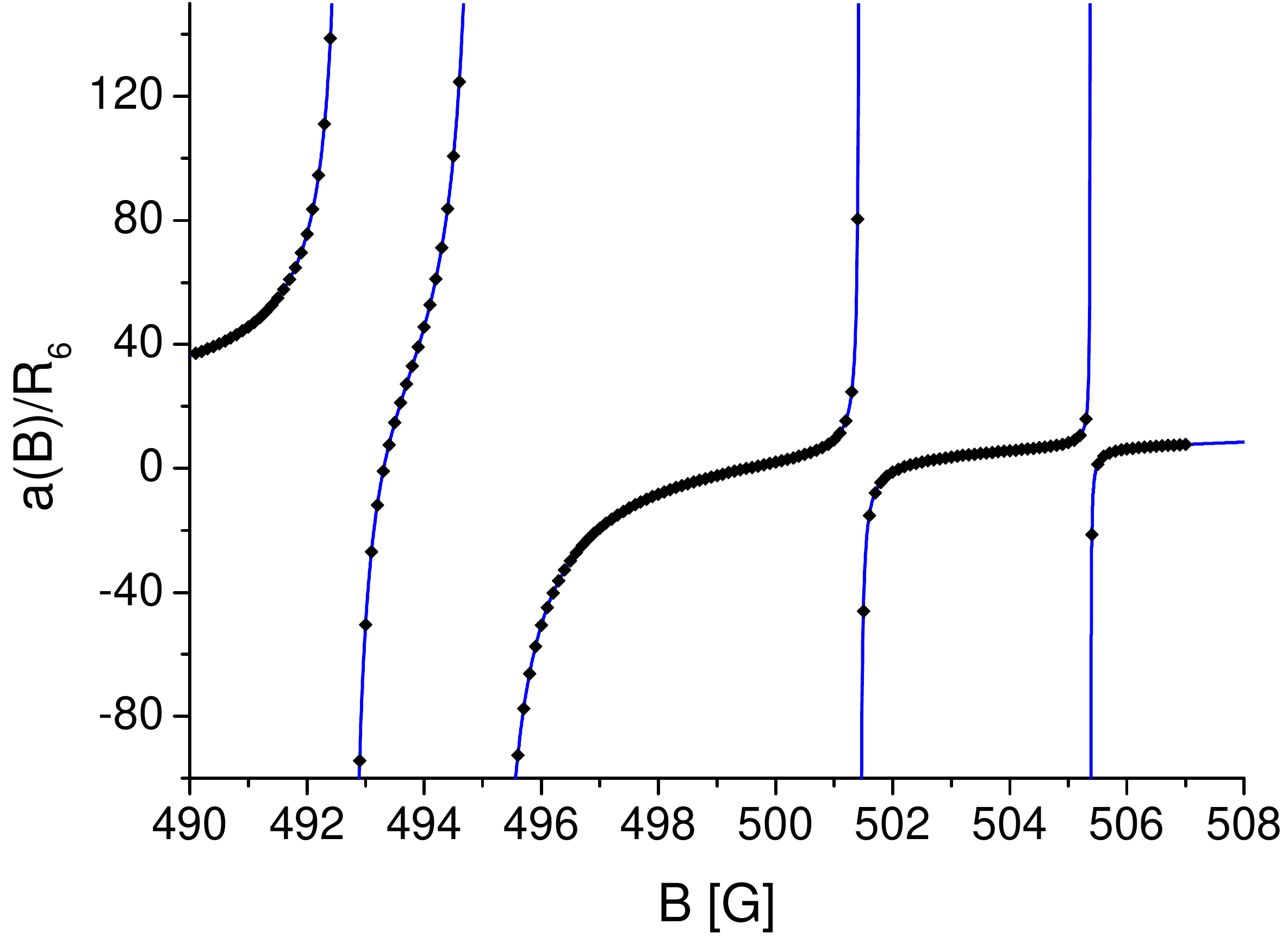}
\includegraphics[width=0.31\linewidth]{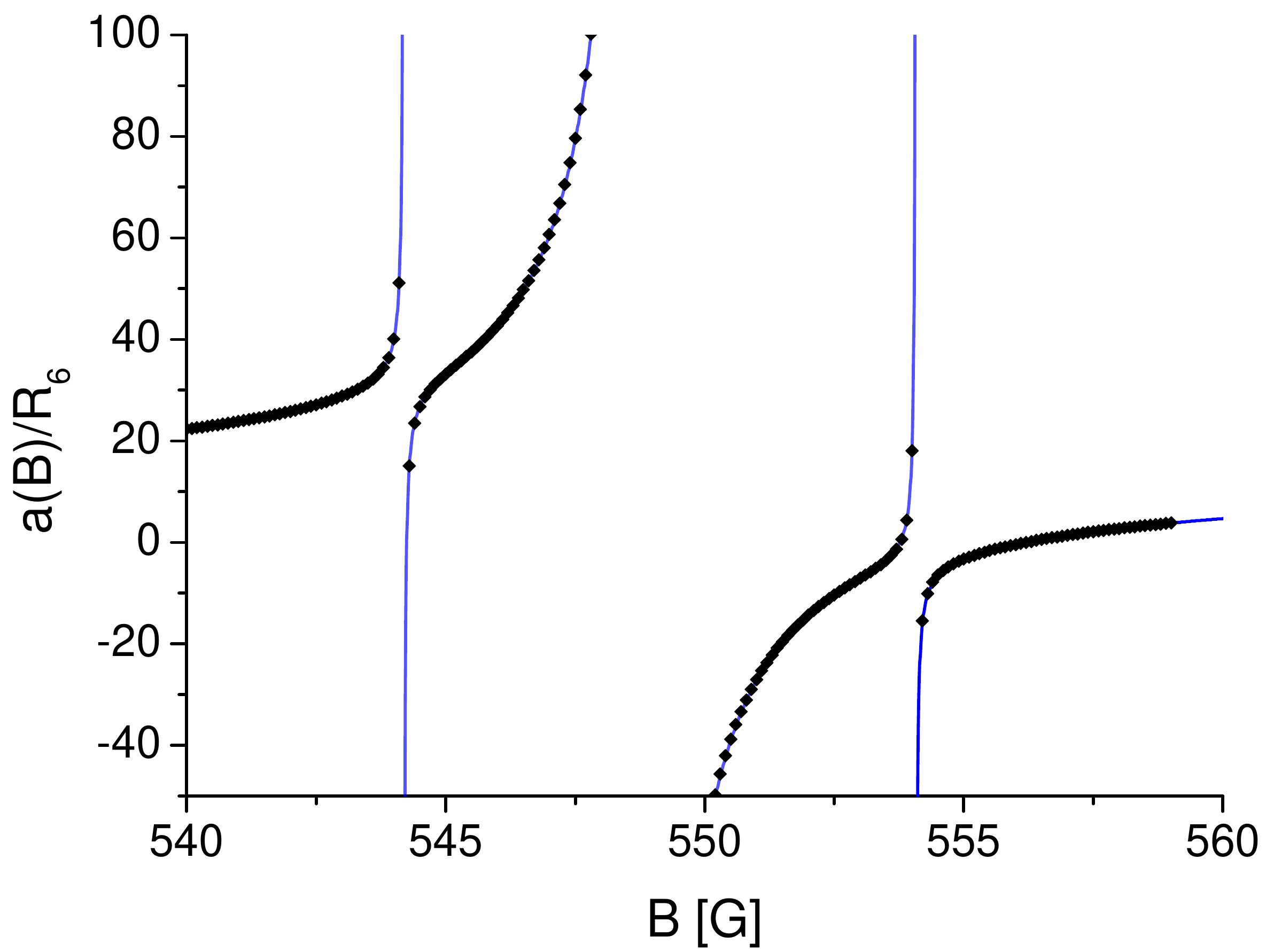}
\includegraphics[width=0.31\linewidth]{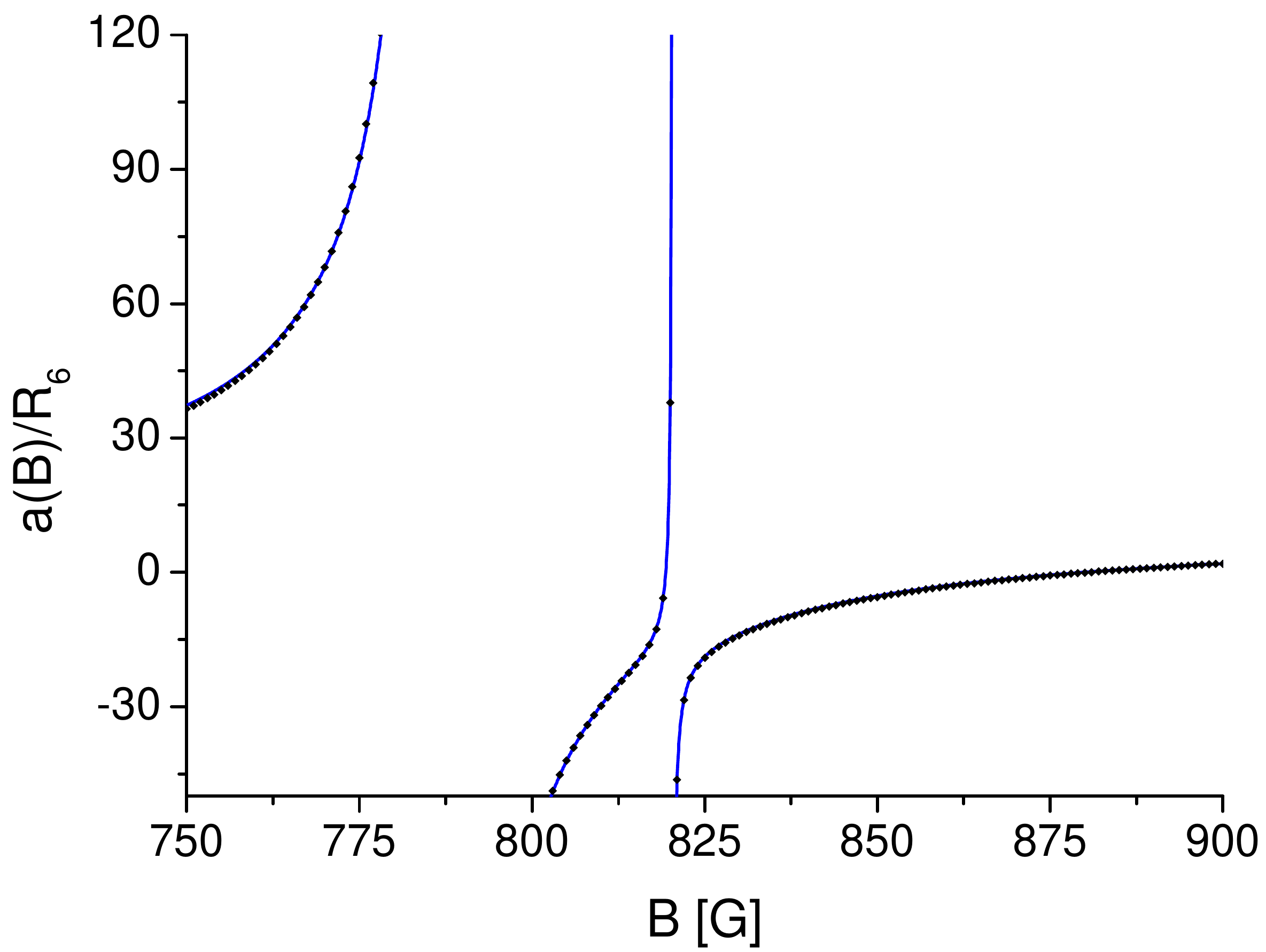}
\caption{\label{cs1}(color online) Some examples of overlapping Feshbach resonances in the collisions of two $f=3,m_f=3$ cesium atoms. Left: a structure of four overlapping $d$-wave Feshbach resonances. Middle: a set of two narrow $g$-wave resonances overlapping with a broad $s$-wave one. Right: a broad $s$-wave Feshbach resonance overlapping with a narrow $d$-wave one. Coupled channel calculations~\cite{Berninger2013} (black points) are compared with the fit basing on formula~\eqref{ascg} (blue lines). The fits are better than $99\%$ accurate in all cases.}
\end{figure*}

Precise Feshbach spectroscopy has been performed for many alkali metal species, and we will use examples of overlapping resonances here that have been studied experimentally for ultracold $f=1,\,m_f=0$  $^7$Li~\cite{Gross2009,Gross2011} and $f=3,\,m_f=3$ cesium~\cite{Chin2004,Berninger2013}.   Both of these species have been used in experimental studies of exotic three-body Efimov physics~\cite{Kraemer2006,Ferlaino2011,Berninger2011,Gross2009,Gross2010,Gross2011}, where it is important to understand the character of the resonances and the mapping of the experimental $B$ field to scattering length.  In order to demonstrate in practice how several overlapping resonances can be described, we apply our MQDT formulas to analyze the results of numerical coupled channels calculations using full Hamiltonian models that have been calibrated to reproduce a variety of experimental data. To account for the changing spin character of the channel states at moderate magnetic fields as well as the influence of resonances which were not included in the fit, we allow a small linear variation with $B$ in $a_{bg}(B) = a_{bg}(1+b(B-B_0))$ as a first-order correction.  We used simple least-square fitting procedures that converge slowly and have to be performed carefully for nonlinear and diverging functions. 

Figures~\ref{li} and \ref{cs1} give the results of our fitting. Figure~\ref{li} shows the two $s$-wave resonances in collision of two $^7$Li atoms in the $f=1,\,m_f=0$ spin channel. For this system the characteristic length $R_6\approx 64.973\,a_0$ ($a_0$ is the Bohr radius) for $C_6 =$ 1393.39 atomic units~\cite{Yan1996} ($E_\mathrm{h} a_0^6$). The coupled channels scattering model was derived from our very accurate model for the collision of $^6$Li atoms~\cite{Zurn2013}.  We fit the binding energy data for the $f=1,\,m_f=+1$ $^7$Li spin channel from Ref.~\cite{Dyke2013}, using a changed scattering length from that of the $^6$Li$_2$ $^1\Sigma_g^+$ potential to account for the failure of isotopic mass scaling, and find coupled channels pole positions at $845.14$ G and $894.00$ G~\cite{GaussNote} for two $f=1$ ,$m_f=1$ atoms, in excellent agreement with the experimental values, 844.9(8) G and 893.7(4) G, reported by Gross {\it et al.}~\cite{Gross2011}.   Our least squares fitting of Eq.~(\ref{ascg}) to a discrete set of coupled channels calculations tabulated on a 1 G grid between 750 G and 950 G yields the same two $B^{res}_i$ positions, with $B_1=857.08$ G, $B_2=828.10$ G, widths $\Delta_1=-73.24$ G, $\Delta_2=-189.74$ G and the background scattering length $a_{bg}/R_6=-0.2339+0.000773(B-800\,G)$ (the definitions ensure that $a_{bg}$ and $\Delta$ have the same sign to fulfill the requirement that $s_{res}$ be positive definite~\cite{Chin2010}). These four $B_i$ and $\Delta_i$ values found are insensitive to the fit range used.  Note that $B_i$ differs in each case from the pole position $B_i^{res}$ because of the large shifts involved. Figure~\ref{li} also shows the relative deviation of the fit from the numerics, indicating that the fit quality is better than $99.5\%$ everywhere. A fit of the same quality is obtained if the product form in Eq.~(\ref{achin}) is used. Using $\delta\mu\approx2.66$MHz$/$G, we obtained the $s_{res}$ parameters for these resonances equal to $0.0467$ for the narrow one and $0.493$ for the wide one. Calculating the $\beta$ parameter defined as in Eq.~\eqref{beta} gives $\beta_1=0.310$ and $\beta_2=1.27$.

In the case of cesium, for which $R_6 = 202.1\,a_0$ and $C_6 =6890.5$ atomic units~\cite{Berninger2013}, many overlapping resonances in various partial waves (up to the $i$-wave) were observed, and several regions of overlapping resonances are present, some quite complex~\cite{Berninger2013}. We focus here on three particular regions of magnetic field for which interesting three-body features have also been reported~\cite{Kraemer2006,Ferlaino2011,Berninger2011}. The first is a set of four $d$-wave resonances near $500$~G, the second involves an $s$-wave and two narrow $g$-wave resonances near $550$~G, and finally the third region has an extremely broad $s$-wave resonance near $800$~G with a narrow $d$-wave one on its shoulder~\cite{note1}. 

Figure~\ref{cs1} shows a comparison of the fits obtained by doing a least-squares fit of Eq.~(\ref{ascg}) to the coupled channels $a(B)$ function from the Supplemental Material for Ref.~\cite{Berninger2013}. Table I lists the fitting parameters found using the respective fitting ranges $490$~G to $508$~G, $530$~G to $570$ G, and $750$ G to $900$ G. Note that the background scattering lengths obtained from locally fitting the resonances in each region differ from the global $a_{bg}$, which is expected to be on the order of $\approx 10\,R_6$. This means that other resonances not included in the fit are near enough to affect the local background and widths. Nevertheless, the fit quality in all cases stays at the $99\%$ level. A comparable quality fit is obtained by fitting to the product form in Eq.~(\ref{achin}).  The resonance at $492.7$~G is especially interesting here, as it is broadened by the presence of two stronger ones, that is, its nearby zero-crossing is shifted $0.63$~G from the pole of the resonance.  This is consistent with the value of $\beta=$ 12.98, where $\beta$ from Eq.~(\ref{beta}) is an analogue of $\alpha$ in Eq.~(\ref{narrow}) for a narrow resonance affected by two broad ones. 

Table I also shows the $s_{res}$ and $\beta$ parameters describing each of the resonances. The $s_{res}$ parameter defined for each resonance is insensitive to fitting ranges.  Our respective values of $s_{res}$ of $160$ and $1480$ for the $554.79$~G and $786.83$ G strong $s$-wave resonances compare with the values $170$ and $1470$ estimated by Chin {\it et al.}~\cite{Chin2010} from an earlier coupled channels model fit to Eq.~(\ref{eq:aB}) for a single isolated resonance. The $\beta$ parameter for the broad resonances tend to be the order of unity.  However, the narrower resonances on their shoulders show large departures of $\beta$ from unity, implying a renormalization of the pole strength due to interactions among the resonances.   We also found that the $554.0$7~G $g$-wave and $820.33$~G $d$-wave resonances, both being narrow ones sitting on the shoulder of the nearby broad $s$-wave one, have respective $s_{res}$ values of 1.6 and 16, differing by an order of magnitude.   These $s_{res}$ values are relevant to explaining the three-body physics that has been explored in the vicinity of these resonances~\cite{Ferlaino2011,Berninger2011,Wang2013}.

When the formula in Eq.~(\ref{narrow}) was applied to each of the narrow resonances near $554$~G and $820$~G, with respective $\alpha$ parameters of $\approx$  $0.35$ and $4.5$, the fit quality was only $90\%$ to $95\%$ accurate.  However, we found that by using a more complex formula that takes into account the small change in the width of the broad resonance due to its interaction with the narrow ones, we were able to obtain a fit accurate to approximately $99\%$, comparable to that obtained with Eqs.~(\ref{ascg}) or \eqref{achin}.

\begin{table*}
\begin{tabular}{|c|c|c|c|c|c|c|c|}
\hline
$B_i^{res} [G] $ & $\ell_i$  & $a_{bg}/R_6$ & $B_i [G] $ & $\Delta_i$ [G] & $\beta_i$ & $\delta\mu_i/E_6$ [G$^{-1}$] & $s_{res}$\\
\hline \hline
492.68 & $d$-wave  & $15.2312(1+0.00368(B-500\,G))$ & 493.315 & 0.1518 & 12.98 & 3.91 & 56\\
495.04 & $d$-wave & $15.2312(1+0.00368(B-500\,G))$ & 499.617 & 3.397 & 1.05 & 3.84 & 100\\
501.44 & $d$-wave & $15.2312(1+0.00368(B-500\,G))$ & 502.186 & 2.088 & 0.093 & 3.79 & 5.4\\
505.38 & $d$-wave & $15.2312(1+0.00368(B-500\,G))$ & 505.483 & 0.2389 & 0.194 & 3.85 & 1.3\\
\hline \hline
544.19 & $g$-wave & $12.7729(1+0.00448(B-550\,G))$ & 544.108 & 0.02179 & 4.95 & 1.52 & 1.0 \\
548.79 & $s$-wave & $12.7729(1+0.00448(B-550\,G))$ & 556.693 & 6.626 & 1.01 & 3.90 & 160 \\
554.07 & $g$-wave & $12.7729(1+0.00448(B-550\,G))$ & 553.793 & 0.5048 & 0.387 & 1.34 & 1.6 \\
\hline \hline
786.17 & $s$-wave & $10.7164(1+0.000469(B-800\,G))$ & 884.638 & 92.26 & 0.856 & 3.66 & 1480 \\
820.32 & $d$-wave & $10.7164(1+0.000469(B-800\,G))$ & 819.386 & 0.4508 & 3.38 & 2.05 & 16\\
\hline
\end{tabular}
\caption{Resonance parameters obtained from fitting to coupled channels calculations for cesium~\cite{Berninger2013} using formula~\eqref{ascg}. The columns show the position of the resonance $B_{res}$ (magnetic field at which the scattering length diverges), its relevant partial wave $\ell$, the background scattering length $a_{bg}$, magnetic field $B_i$ at which the bare bound state crosses the threshold, resonance width $\Delta_i$, ratio $\beta_i$ from Eq.~\eqref{beta}, difference of magnetic moments $\delta\mu_i$  (from the coupled channel calculations), and an estimate for $s_{res}\approx0.478\beta_i a_{bg}\Delta_i \delta\mu_i$.}
\end{table*}

\section{Conclusion}

In conclusion, we presented a simple analytical model to describe the variation of the $s$-wave scattering length with magnetic field when there is an arbitrary number of Feshbach resonances. We introduced simple parameters characterizing the system, analogous to the resonance width and background scattering length in the two-channel case. We discussed the non-separability of the scattering length formula for overlapping resonances and provided examples where we accurately reproduced coupled channels numerical results and found the resonance parameters. Apart from characterization of overlapping sets of resonances, our model should be quite helpful for precise mapping of the scattering length to the laboratory $B$ field, which is a critical aspect in the interpretation of experiments with three-body recombination and Efimov physics.

This work was supported by the Foundation for Polish Science International PhD Projects Programme co-financed by the EU European Regional Development Fund, by AFOSR MURI Grant No. FA9550-09-1-0617, and in part by the National Science Foundation under Grant No. NSF PHY11-25915.

\bibliography{Allrefs}

\begin{thebibliography}{49}
\expandafter\ifx\csname natexlab\endcsname\relax\def\natexlab#1{#1}\fi
\expandafter\ifx\csname bibnamefont\endcsname\relax
  \def\bibnamefont#1{#1}\fi
\expandafter\ifx\csname bibfnamefont\endcsname\relax
  \def\bibfnamefont#1{#1}\fi
\expandafter\ifx\csname citenamefont\endcsname\relax
  \def\citenamefont#1{#1}\fi
\expandafter\ifx\csname url\endcsname\relax
  \def\url#1{\texttt{#1}}\fi
\expandafter\ifx\csname urlprefix\endcsname\relax\def\urlprefix{URL }\fi
\providecommand{\bibinfo}[2]{#2}
\providecommand{\eprint}[2][]{\url{#2}}

\bibitem[{\citenamefont{Chin et~al.}(2010)\citenamefont{Chin, Grimm, Julienne,
  and Tiesinga}}]{Chin2010}
\bibinfo{author}{\bibfnamefont{C.}~\bibnamefont{Chin}},
  \bibinfo{author}{\bibfnamefont{R.}~\bibnamefont{Grimm}},
  \bibinfo{author}{\bibfnamefont{P.~S.} \bibnamefont{Julienne}},
  \bibnamefont{and} \bibinfo{author}{\bibfnamefont{E.}~\bibnamefont{Tiesinga}},
  \bibinfo{journal}{Rev. Mod. Phys.} \textbf{\bibinfo{volume}{82}},
  \bibinfo{pages}{1225} (\bibinfo{year}{2010}).

\bibitem[{\citenamefont{Inouye et~al.}(1998)\citenamefont{Inouye, Andrews,
  Stenger, Miesner, Stamper-Kurn, and Ketterle}}]{Inouye1998}
\bibinfo{author}{\bibfnamefont{S.}~\bibnamefont{Inouye}},
  \bibinfo{author}{\bibfnamefont{M.~R.} \bibnamefont{Andrews}},
  \bibinfo{author}{\bibfnamefont{J.}~\bibnamefont{Stenger}},
  \bibinfo{author}{\bibfnamefont{H.-J.} \bibnamefont{Miesner}},
  \bibinfo{author}{\bibfnamefont{D.~M.} \bibnamefont{Stamper-Kurn}},
  \bibnamefont{and} \bibinfo{author}{\bibfnamefont{W.}~\bibnamefont{Ketterle}},
  \bibinfo{journal}{Nature} \textbf{\bibinfo{volume}{392}},
  \bibinfo{pages}{151} (\bibinfo{year}{1998}).

\bibitem[{\citenamefont{Cornish et~al.}(2000)\citenamefont{Cornish, Claussen,
  Roberts, Cornell, and Wieman}}]{Cornish2000}
\bibinfo{author}{\bibfnamefont{S.~L.} \bibnamefont{Cornish}},
  \bibinfo{author}{\bibfnamefont{N.~R.} \bibnamefont{Claussen}},
  \bibinfo{author}{\bibfnamefont{J.~L.} \bibnamefont{Roberts}},
  \bibinfo{author}{\bibfnamefont{E.~A.} \bibnamefont{Cornell}},
  \bibnamefont{and} \bibinfo{author}{\bibfnamefont{C.~E.}
  \bibnamefont{Wieman}}, \bibinfo{journal}{Phys. Rev. Lett.}
  \textbf{\bibinfo{volume}{85}}, \bibinfo{pages}{1795} (\bibinfo{year}{2000}).

\bibitem[{\citenamefont{Bourdel et~al.}(2004)\citenamefont{Bourdel, Khaykovich,
  Cubizolles, Zhang, Chevy, Teichmann, Tarruell, Kokkelmans, and
  Salomon}}]{Bourdel2004}
\bibinfo{author}{\bibfnamefont{T.}~\bibnamefont{Bourdel}},
  \bibinfo{author}{\bibfnamefont{L.}~\bibnamefont{Khaykovich}},
  \bibinfo{author}{\bibfnamefont{J.}~\bibnamefont{Cubizolles}},
  \bibinfo{author}{\bibfnamefont{J.}~\bibnamefont{Zhang}},
  \bibinfo{author}{\bibfnamefont{F.}~\bibnamefont{Chevy}},
  \bibinfo{author}{\bibfnamefont{M.}~\bibnamefont{Teichmann}},
  \bibinfo{author}{\bibfnamefont{L.}~\bibnamefont{Tarruell}},
  \bibinfo{author}{\bibfnamefont{S.~J. J. M.~F.} \bibnamefont{Kokkelmans}},
  \bibnamefont{and} \bibinfo{author}{\bibfnamefont{C.}~\bibnamefont{Salomon}},
  \bibinfo{journal}{Phys. Rev. Lett.} \textbf{\bibinfo{volume}{93}},
  \bibinfo{pages}{050401} (\bibinfo{year}{2004}).

\bibitem[{\citenamefont{Zwierlein et~al.}(2004)\citenamefont{Zwierlein, Stan,
  Schunck, Raupach, Kerman, and Ketterle}}]{Zwierlein2004}
\bibinfo{author}{\bibfnamefont{M.~W.} \bibnamefont{Zwierlein}},
  \bibinfo{author}{\bibfnamefont{C.~A.} \bibnamefont{Stan}},
  \bibinfo{author}{\bibfnamefont{C.~H.} \bibnamefont{Schunck}},
  \bibinfo{author}{\bibfnamefont{S.~M.~F.} \bibnamefont{Raupach}},
  \bibinfo{author}{\bibfnamefont{A.~J.} \bibnamefont{Kerman}},
  \bibnamefont{and} \bibinfo{author}{\bibfnamefont{W.}~\bibnamefont{Ketterle}},
  \bibinfo{journal}{Phys. Rev. Lett.} \textbf{\bibinfo{volume}{92}},
  \bibinfo{pages}{120403} (\bibinfo{year}{2004}).

\bibitem[{\citenamefont{Jochim et~al.}(2003)\citenamefont{Jochim, Bartenstein,
  Altmeyer, Hendl, Riedl, Chin, {Hecker Denschlag}, and Grimm}}]{Jochim2003a}
\bibinfo{author}{\bibfnamefont{S.}~\bibnamefont{Jochim}},
  \bibinfo{author}{\bibfnamefont{M.}~\bibnamefont{Bartenstein}},
  \bibinfo{author}{\bibfnamefont{A.}~\bibnamefont{Altmeyer}},
  \bibinfo{author}{\bibfnamefont{G.}~\bibnamefont{Hendl}},
  \bibinfo{author}{\bibfnamefont{S.}~\bibnamefont{Riedl}},
  \bibinfo{author}{\bibfnamefont{C.}~\bibnamefont{Chin}},
  \bibinfo{author}{\bibfnamefont{J.}~\bibnamefont{{Hecker Denschlag}}},
  \bibnamefont{and} \bibinfo{author}{\bibfnamefont{R.}~\bibnamefont{Grimm}},
  \bibinfo{journal}{Science} \textbf{\bibinfo{volume}{302}},
  \bibinfo{pages}{2101} (\bibinfo{year}{2003}).

\bibitem[{\citenamefont{Lang et~al.}(2008)\citenamefont{Lang, Winkler, Strauss,
  Grimm, and {Hecker Denschlag}}}]{Lang2008}
\bibinfo{author}{\bibfnamefont{F.}~\bibnamefont{Lang}},
  \bibinfo{author}{\bibfnamefont{K.}~\bibnamefont{Winkler}},
  \bibinfo{author}{\bibfnamefont{C.}~\bibnamefont{Strauss}},
  \bibinfo{author}{\bibfnamefont{R.}~\bibnamefont{Grimm}}, \bibnamefont{and}
  \bibinfo{author}{\bibfnamefont{J.}~\bibnamefont{{Hecker Denschlag}}},
  \bibinfo{journal}{Phys. Rev. Lett.} \textbf{\bibinfo{volume}{101}},
  \bibinfo{pages}{133005} (\bibinfo{year}{2008}).

\bibitem[{\citenamefont{Moerdijk et~al.}(1995)\citenamefont{Moerdijk, Verhaar,
  and Axelsson}}]{Moerdijk1995}
\bibinfo{author}{\bibfnamefont{A.~J.} \bibnamefont{Moerdijk}},
  \bibinfo{author}{\bibfnamefont{B.~J.} \bibnamefont{Verhaar}},
  \bibnamefont{and} \bibinfo{author}{\bibfnamefont{A.}~\bibnamefont{Axelsson}},
  \bibinfo{journal}{Phys. Rev. A} \textbf{\bibinfo{volume}{51}},
  \bibinfo{pages}{4852} (\bibinfo{year}{1995}).

\bibitem[{\citenamefont{Timmermans et~al.}(1999)\citenamefont{Timmermans,
  Tommasini, Hussein, and Kerman}}]{Timmermans1999}
\bibinfo{author}{\bibfnamefont{E.}~\bibnamefont{Timmermans}},
  \bibinfo{author}{\bibfnamefont{P.}~\bibnamefont{Tommasini}},
  \bibinfo{author}{\bibfnamefont{M.}~\bibnamefont{Hussein}}, \bibnamefont{and}
  \bibinfo{author}{\bibfnamefont{A.}~\bibnamefont{Kerman}},
  \bibinfo{journal}{Phys. Rep.} \textbf{\bibinfo{volume}{315}},
  \bibinfo{pages}{199} (\bibinfo{year}{1999}).

\bibitem[{\citenamefont{Kraemer et~al.}(2006)\citenamefont{Kraemer, Mark,
  Waldburger, Danzl, Chin, Engeser, Lange, Pilch, Jaakkola, N\"agerl
  et~al.}}]{Kraemer2006}
\bibinfo{author}{\bibfnamefont{T.}~\bibnamefont{Kraemer}},
  \bibinfo{author}{\bibfnamefont{M.}~\bibnamefont{Mark}},
  \bibinfo{author}{\bibfnamefont{P.}~\bibnamefont{Waldburger}},
  \bibinfo{author}{\bibfnamefont{J.~G.} \bibnamefont{Danzl}},
  \bibinfo{author}{\bibfnamefont{C.}~\bibnamefont{Chin}},
  \bibinfo{author}{\bibfnamefont{B.}~\bibnamefont{Engeser}},
  \bibinfo{author}{\bibfnamefont{A.~D.} \bibnamefont{Lange}},
  \bibinfo{author}{\bibfnamefont{K.}~\bibnamefont{Pilch}},
  \bibinfo{author}{\bibfnamefont{A.}~\bibnamefont{Jaakkola}},
  \bibinfo{author}{\bibfnamefont{H.-C.} \bibnamefont{N\"agerl}},
  \bibnamefont{et~al.}, \bibinfo{journal}{Nature}
  \textbf{\bibinfo{volume}{440}}, \bibinfo{pages}{315} (\bibinfo{year}{2006}).

\bibitem[{\citenamefont{Knoop et~al.}(2009)\citenamefont{Knoop, Ferlaino, Mark,
  Danzl, Kraemer, N\"agerl, and Grimm}}]{Knoop2008}
\bibinfo{author}{\bibfnamefont{S.}~\bibnamefont{Knoop}},
  \bibinfo{author}{\bibfnamefont{F.}~\bibnamefont{Ferlaino}},
  \bibinfo{author}{\bibfnamefont{M.}~\bibnamefont{Mark}},
  \bibinfo{author}{\bibfnamefont{J.~G.} \bibnamefont{Danzl}},
  \bibinfo{author}{\bibfnamefont{T.}~\bibnamefont{Kraemer}},
  \bibinfo{author}{\bibfnamefont{H.-C.} \bibnamefont{N\"agerl}},
  \bibnamefont{and} \bibinfo{author}{\bibfnamefont{R.}~\bibnamefont{Grimm}},
  \bibinfo{journal}{Nature Phys.} \textbf{\bibinfo{volume}{5}},
  \bibinfo{pages}{227} (\bibinfo{year}{2009}).

\bibitem[{\citenamefont{Zaccanti et~al.}(2009)\citenamefont{Zaccanti, Deissler,
  D’Errico, Fattori, Jona-Lasinio, M{\"u}ller, Roati, Inguscio, and
  Modugno}}]{Zaccanti2009}
\bibinfo{author}{\bibfnamefont{M.}~\bibnamefont{Zaccanti}},
  \bibinfo{author}{\bibfnamefont{B.}~\bibnamefont{Deissler}},
  \bibinfo{author}{\bibfnamefont{C.}~\bibnamefont{D’Errico}},
  \bibinfo{author}{\bibfnamefont{M.}~\bibnamefont{Fattori}},
  \bibinfo{author}{\bibfnamefont{M.}~\bibnamefont{Jona-Lasinio}},
  \bibinfo{author}{\bibfnamefont{S.}~\bibnamefont{M{\"u}ller}},
  \bibinfo{author}{\bibfnamefont{G.}~\bibnamefont{Roati}},
  \bibinfo{author}{\bibfnamefont{M.}~\bibnamefont{Inguscio}}, \bibnamefont{and}
  \bibinfo{author}{\bibfnamefont{G.}~\bibnamefont{Modugno}},
  \bibinfo{journal}{Nature Physics} \textbf{\bibinfo{volume}{5}},
  \bibinfo{pages}{586} (\bibinfo{year}{2009}).

\bibitem[{\citenamefont{Gross et~al.}(2010{\natexlab{a}})\citenamefont{Gross,
  Shotan, Kokkelmans, and Khaykovich}}]{Khaykovich2010}
\bibinfo{author}{\bibfnamefont{N.}~\bibnamefont{Gross}},
  \bibinfo{author}{\bibfnamefont{Z.}~\bibnamefont{Shotan}},
  \bibinfo{author}{\bibfnamefont{S.}~\bibnamefont{Kokkelmans}},
  \bibnamefont{and}
  \bibinfo{author}{\bibfnamefont{L.}~\bibnamefont{Khaykovich}},
  \bibinfo{journal}{Phys. Rev. Lett.} \textbf{\bibinfo{volume}{105}},
  \bibinfo{pages}{103203} (\bibinfo{year}{2010}{\natexlab{a}}).

\bibitem[{\citenamefont{Ferlaino et~al.}(2011)\citenamefont{Ferlaino, Zenesini,
  Berninger, Huang, Naegerl, and Grimm}}]{Ferlaino2011}
\bibinfo{author}{\bibfnamefont{F.}~\bibnamefont{Ferlaino}},
  \bibinfo{author}{\bibfnamefont{A.}~\bibnamefont{Zenesini}},
  \bibinfo{author}{\bibfnamefont{M.}~\bibnamefont{Berninger}},
  \bibinfo{author}{\bibfnamefont{B.}~\bibnamefont{Huang}},
  \bibinfo{author}{\bibfnamefont{H.~C.} \bibnamefont{Naegerl}},
  \bibnamefont{and} \bibinfo{author}{\bibfnamefont{R.}~\bibnamefont{Grimm}},
  \bibinfo{journal}{Few Body Sys.} \textbf{\bibinfo{volume}{51}},
  \bibinfo{pages}{113} (\bibinfo{year}{2011}).

\bibitem[{\citenamefont{Berninger et~al.}(2011)\citenamefont{Berninger,
  Zenesini, Huang, Harm, N\"agerl, Ferlaino, Grimm, Julienne, and
  Hutson}}]{Berninger2011}
\bibinfo{author}{\bibfnamefont{M.}~\bibnamefont{Berninger}},
  \bibinfo{author}{\bibfnamefont{A.}~\bibnamefont{Zenesini}},
  \bibinfo{author}{\bibfnamefont{B.}~\bibnamefont{Huang}},
  \bibinfo{author}{\bibfnamefont{W.}~\bibnamefont{Harm}},
  \bibinfo{author}{\bibfnamefont{H.-C.} \bibnamefont{N\"agerl}},
  \bibinfo{author}{\bibfnamefont{F.}~\bibnamefont{Ferlaino}},
  \bibinfo{author}{\bibfnamefont{R.}~\bibnamefont{Grimm}},
  \bibinfo{author}{\bibfnamefont{P.~S.} \bibnamefont{Julienne}},
  \bibnamefont{and} \bibinfo{author}{\bibfnamefont{J.~M.}
  \bibnamefont{Hutson}}, \bibinfo{journal}{Phys. Rev. Lett.}
  \textbf{\bibinfo{volume}{107}}, \bibinfo{pages}{120401}
  (\bibinfo{year}{2011}).

\bibitem[{\citenamefont{Berninger et~al.}(2013)\citenamefont{Berninger,
  Zenesini, Huang, Harm, N\"agerl, Ferlaino, Grimm, Julienne, and
  Hutson}}]{Berninger2013}
\bibinfo{author}{\bibfnamefont{M.}~\bibnamefont{Berninger}},
  \bibinfo{author}{\bibfnamefont{A.}~\bibnamefont{Zenesini}},
  \bibinfo{author}{\bibfnamefont{B.}~\bibnamefont{Huang}},
  \bibinfo{author}{\bibfnamefont{W.}~\bibnamefont{Harm}},
  \bibinfo{author}{\bibfnamefont{H.-C.} \bibnamefont{N\"agerl}},
  \bibinfo{author}{\bibfnamefont{F.}~\bibnamefont{Ferlaino}},
  \bibinfo{author}{\bibfnamefont{R.}~\bibnamefont{Grimm}},
  \bibinfo{author}{\bibfnamefont{P.~S.} \bibnamefont{Julienne}},
  \bibnamefont{and} \bibinfo{author}{\bibfnamefont{J.~M.}
  \bibnamefont{Hutson}}, \bibinfo{journal}{Phys. Rev. A}
  \textbf{\bibinfo{volume}{87}}, \bibinfo{pages}{032517}
  (\bibinfo{year}{2013}).

\bibitem[{\citenamefont{Dyke et~al.}(2013)\citenamefont{Dyke, Pollack, and
  Hulet}}]{Dyke2013}
\bibinfo{author}{\bibfnamefont{P.}~\bibnamefont{Dyke}},
  \bibinfo{author}{\bibfnamefont{S.~E.} \bibnamefont{Pollack}},
  \bibnamefont{and} \bibinfo{author}{\bibfnamefont{R.~G.} \bibnamefont{Hulet}},
  \bibinfo{journal}{arXiv:1302.0281}  (\bibinfo{year}{2013}).

\bibitem[{\citenamefont{Mies}(1984)}]{Mies1984a}
\bibinfo{author}{\bibfnamefont{F.~H.} \bibnamefont{Mies}}, \bibinfo{journal}{J.
  Chem. Phys.} \textbf{\bibinfo{volume}{80}}, \bibinfo{pages}{2514}
  (\bibinfo{year}{1984}).

\bibitem[{\citenamefont{Mies and Julienne}(1984)}]{Mies1984b}
\bibinfo{author}{\bibfnamefont{F.~H.} \bibnamefont{Mies}} \bibnamefont{and}
  \bibinfo{author}{\bibfnamefont{P.~S.} \bibnamefont{Julienne}},
  \bibinfo{journal}{J. Chem. Phys.} \textbf{\bibinfo{volume}{80}},
  \bibinfo{pages}{2526} (\bibinfo{year}{1984}).

\bibitem[{\citenamefont{Julienne and Mies}(1989)}]{Julienne1989}
\bibinfo{author}{\bibfnamefont{P.~S.} \bibnamefont{Julienne}} \bibnamefont{and}
  \bibinfo{author}{\bibfnamefont{F.~H.} \bibnamefont{Mies}},
  \bibinfo{journal}{J. Opt. Soc. Am. B} \textbf{\bibinfo{volume}{6}},
  \bibinfo{pages}{2257} (\bibinfo{year}{1989}).

\bibitem[{\citenamefont{Mies and Raoult}(2000)}]{Mies2000}
\bibinfo{author}{\bibfnamefont{F.~H.} \bibnamefont{Mies}} \bibnamefont{and}
  \bibinfo{author}{\bibfnamefont{M.}~\bibnamefont{Raoult}},
  \bibinfo{journal}{Phys. Rev. A} \textbf{\bibinfo{volume}{62}},
  \bibinfo{pages}{012708} (\bibinfo{year}{2000}).

\bibitem[{\citenamefont{Julienne and Gao}(2006)}]{Julienne2006}
\bibinfo{author}{\bibfnamefont{P.~S.} \bibnamefont{Julienne}} \bibnamefont{and}
  \bibinfo{author}{\bibfnamefont{B.}~\bibnamefont{Gao}}, in
  \emph{\bibinfo{booktitle}{Atomic Physics 20}}, edited by
  \bibinfo{editor}{\bibfnamefont{C.}~\bibnamefont{Roos}},
  \bibinfo{editor}{\bibfnamefont{H.}~\bibnamefont{H{\"a}ffner}},
  \bibnamefont{and} \bibinfo{editor}{\bibfnamefont{R.}~\bibnamefont{Blatt}}
  (\bibinfo{publisher}{AIP, Melville, New York}, \bibinfo{year}{2006}), pp.
  \bibinfo{pages}{261--268}.

\bibitem[{\citenamefont{Julienne}(2009{\natexlab{a}})}]{Julienne2009b}
\bibinfo{author}{\bibfnamefont{P.~S.} \bibnamefont{Julienne}},
  \bibinfo{journal}{Chapter 6 of {\it Cold Molecules: Theory, Experiment,
  Applications}, ed. by R. V. Krems, W. C. Stwalley, B. Friedrich, CRC Press
  (arXiv:0902.1727)} pp. \bibinfo{pages}{221--243}
  (\bibinfo{year}{2009}{\natexlab{a}}).

\bibitem[{\citenamefont{Lange et~al.}(2009)\citenamefont{Lange, Pilch,
  Prantner, Ferlaino, Engeser, N\"agerl, Grimm, and Chin}}]{Lange2009}
\bibinfo{author}{\bibfnamefont{A.~D.} \bibnamefont{Lange}},
  \bibinfo{author}{\bibfnamefont{K.}~\bibnamefont{Pilch}},
  \bibinfo{author}{\bibfnamefont{A.}~\bibnamefont{Prantner}},
  \bibinfo{author}{\bibfnamefont{F.}~\bibnamefont{Ferlaino}},
  \bibinfo{author}{\bibfnamefont{B.}~\bibnamefont{Engeser}},
  \bibinfo{author}{\bibfnamefont{H.-C.} \bibnamefont{N\"agerl}},
  \bibinfo{author}{\bibfnamefont{R.}~\bibnamefont{Grimm}}, \bibnamefont{and}
  \bibinfo{author}{\bibfnamefont{C.}~\bibnamefont{Chin}},
  \bibinfo{journal}{Phys. Rev. A} \textbf{\bibinfo{volume}{79}},
  \bibinfo{pages}{013622} (\bibinfo{year}{2009}).

\bibitem[{\citenamefont{K\"ohler et~al.}(2006)\citenamefont{K\"ohler,
  {G}\'oral, and {J}ulienne}}]{Kohler2006}
\bibinfo{author}{\bibfnamefont{T.}~\bibnamefont{K\"ohler}},
  \bibinfo{author}{\bibfnamefont{K.}~\bibnamefont{{G}\'oral}},
  \bibnamefont{and} \bibinfo{author}{\bibfnamefont{P.~S.}
  \bibnamefont{{J}ulienne}}, \bibinfo{journal}{Rev. {M}od. {P}hys.}
  \textbf{\bibinfo{volume}{78}}, \bibinfo{pages}{1311} (\bibinfo{year}{2006}).

\bibitem[{\citenamefont{Burke et~al.}(1998)\citenamefont{Burke, Greene, and
  Bohn}}]{Burke1998}
\bibinfo{author}{\bibfnamefont{J.~P.} \bibnamefont{Burke}},
  \bibinfo{author}{\bibfnamefont{C.~H.} \bibnamefont{Greene}},
  \bibnamefont{and} \bibinfo{author}{\bibfnamefont{J.~L.} \bibnamefont{Bohn}},
  \bibinfo{journal}{Phys. Rev. Lett.} \textbf{\bibinfo{volume}{81}},
  \bibinfo{pages}{3355} (\bibinfo{year}{1998}).

\bibitem[{\citenamefont{Gao}(2008)}]{Gao2008}
\bibinfo{author}{\bibfnamefont{B.}~\bibnamefont{Gao}}, \bibinfo{journal}{Phys.
  Rev. A} \textbf{\bibinfo{volume}{78}}, \bibinfo{pages}{012702}
  (\bibinfo{year}{2008}).

\bibitem[{\citenamefont{Gao}(2011)}]{Gao2011}
\bibinfo{author}{\bibfnamefont{B.}~\bibnamefont{Gao}}, \bibinfo{journal}{Phys.
  Rev. A} \textbf{\bibinfo{volume}{83}}, \bibinfo{pages}{062712}
  (\bibinfo{year}{2011}).

\bibitem[{\citenamefont{Idziaszek et~al.}(2011)\citenamefont{Idziaszek, Simoni,
  Calarco, and Julienne}}]{NJP2011}
\bibinfo{author}{\bibfnamefont{Z.}~\bibnamefont{Idziaszek}},
  \bibinfo{author}{\bibfnamefont{A.}~\bibnamefont{Simoni}},
  \bibinfo{author}{\bibfnamefont{T.}~\bibnamefont{Calarco}}, \bibnamefont{and}
  \bibinfo{author}{\bibfnamefont{P.~S.} \bibnamefont{Julienne}},
  \bibinfo{journal}{New Journal of Physics} \textbf{\bibinfo{volume}{13}},
  \bibinfo{pages}{083005} (\bibinfo{year}{2011}).

\bibitem[{\citenamefont{Gao}(1998)}]{Gao1998a}
\bibinfo{author}{\bibfnamefont{B.}~\bibnamefont{Gao}}, \bibinfo{journal}{Phys.
  Rev. A} \textbf{\bibinfo{volume}{58}}, \bibinfo{pages}{1728}
  (\bibinfo{year}{1998}).

\bibitem[{\citenamefont{Gao}(2000)}]{Gao2000}
\bibinfo{author}{\bibfnamefont{B.}~\bibnamefont{Gao}}, \bibinfo{journal}{Phys.
  Rev. A} \textbf{\bibinfo{volume}{62}}, \bibinfo{pages}{050702}
  (\bibinfo{year}{2000}).

\bibitem[{\citenamefont{Mies et~al.}(2000)\citenamefont{Mies, {T}iesinga, and
  {J}ulienne}}]{Mies2000a}
\bibinfo{author}{\bibfnamefont{F.~H.} \bibnamefont{Mies}},
  \bibinfo{author}{\bibfnamefont{E.}~\bibnamefont{{T}iesinga}},
  \bibnamefont{and} \bibinfo{author}{\bibfnamefont{P.~S.}
  \bibnamefont{{J}ulienne}}, \bibinfo{journal}{Phys. {R}ev. {A}}
  \textbf{\bibinfo{volume}{61}}, \bibinfo{pages}{022721}
  (\bibinfo{year}{2000}).

\bibitem[{\citenamefont{Nygaard et~al.}(2006)\citenamefont{Nygaard, Schneider,
  and Julienne}}]{Nygaard2006}
\bibinfo{author}{\bibfnamefont{N.}~\bibnamefont{Nygaard}},
  \bibinfo{author}{\bibfnamefont{B.~I.} \bibnamefont{Schneider}},
  \bibnamefont{and} \bibinfo{author}{\bibfnamefont{P.~S.}
  \bibnamefont{Julienne}}, \bibinfo{journal}{Phys. Rev. A}
  \textbf{\bibinfo{volume}{73}}, \bibinfo{pages}{042705}
  (\bibinfo{year}{2006}).

\bibitem[{\citenamefont{Gribakin and Flambaum}(1993)}]{Gribakin1993}
\bibinfo{author}{\bibfnamefont{G.~F.} \bibnamefont{Gribakin}} \bibnamefont{and}
  \bibinfo{author}{\bibfnamefont{V.~V.} \bibnamefont{Flambaum}},
  \bibinfo{journal}{Phys. Rev. A} \textbf{\bibinfo{volume}{48}},
  \bibinfo{pages}{546} (\bibinfo{year}{1993}).

\bibitem[{\citenamefont{Julienne}(2009{\natexlab{b}})}]{Julienne2009}
\bibinfo{author}{\bibfnamefont{P.~S.} \bibnamefont{Julienne}},
  \bibinfo{journal}{Faraday Discuss.} \textbf{\bibinfo{volume}{142}},
  \bibinfo{pages}{361} (\bibinfo{year}{2009}{\natexlab{b}}).

\bibitem[{\citenamefont{Bolda et~al.}(2002)\citenamefont{Bolda, Tiesinga, and
  Julienne}}]{Bolda2002}
\bibinfo{author}{\bibfnamefont{E.~L.} \bibnamefont{Bolda}},
  \bibinfo{author}{\bibfnamefont{E.}~\bibnamefont{Tiesinga}}, \bibnamefont{and}
  \bibinfo{author}{\bibfnamefont{P.~S.} \bibnamefont{Julienne}},
  \bibinfo{journal}{Phys. Rev. A} \textbf{\bibinfo{volume}{66}},
  \bibinfo{pages}{013403} (\bibinfo{year}{2002}).

\bibitem[{\citenamefont{Krych and Idziaszek}(2009)}]{Krych2009}
\bibinfo{author}{\bibfnamefont{M.}~\bibnamefont{Krych}} \bibnamefont{and}
  \bibinfo{author}{\bibfnamefont{Z.}~\bibnamefont{Idziaszek}},
  \bibinfo{journal}{Phys. Rev. A} \textbf{\bibinfo{volume}{80}},
  \bibinfo{pages}{022710} (\bibinfo{year}{2009}).

\bibitem[{\citenamefont{Idziaszek and Julienne}(2010)}]{Idziaszek2010}
\bibinfo{author}{\bibfnamefont{Z.}~\bibnamefont{Idziaszek}} \bibnamefont{and}
  \bibinfo{author}{\bibfnamefont{P.~S.} \bibnamefont{Julienne}},
  \bibinfo{journal}{Phys. Rev. Lett.} \textbf{\bibinfo{volume}{104}},
  \bibinfo{pages}{113202} (\bibinfo{year}{2010}).

\bibitem[{\citenamefont{Wang and Julienne}(2013)}]{Wang2013}
\bibinfo{author}{\bibfnamefont{Y.}~\bibnamefont{Wang}} \bibnamefont{and}
  \bibinfo{author}{\bibfnamefont{P.~S.} \bibnamefont{Julienne}},
  \bibinfo{journal}{unpublished}  (\bibinfo{year}{2013}).

\bibitem[{\citenamefont{Takekoshi et~al.}(2012)\citenamefont{Takekoshi,
  Debatin, Rameshan, Ferlaino, Grimm, N\"agerl, Le~Sueur, Hutson, Julienne,
  Kotochigova et~al.}}]{Takekoshi2012}
\bibinfo{author}{\bibfnamefont{T.}~\bibnamefont{Takekoshi}},
  \bibinfo{author}{\bibfnamefont{M.}~\bibnamefont{Debatin}},
  \bibinfo{author}{\bibfnamefont{R.}~\bibnamefont{Rameshan}},
  \bibinfo{author}{\bibfnamefont{F.}~\bibnamefont{Ferlaino}},
  \bibinfo{author}{\bibfnamefont{R.}~\bibnamefont{Grimm}},
  \bibinfo{author}{\bibfnamefont{H.-C.} \bibnamefont{N\"agerl}},
  \bibinfo{author}{\bibfnamefont{C.~R.} \bibnamefont{Le~Sueur}},
  \bibinfo{author}{\bibfnamefont{J.~M.} \bibnamefont{Hutson}},
  \bibinfo{author}{\bibfnamefont{P.~S.} \bibnamefont{Julienne}},
  \bibinfo{author}{\bibfnamefont{S.}~\bibnamefont{Kotochigova}},
  \bibnamefont{et~al.}, \bibinfo{journal}{Phys. Rev. A}
  \textbf{\bibinfo{volume}{85}}, \bibinfo{pages}{032506}
  (\bibinfo{year}{2012}).

\bibitem[{\citenamefont{Gross et~al.}(2011)\citenamefont{Gross, Shotan,
  Machtey, Kokkelmans, and Khaykovich}}]{Gross2011}
\bibinfo{author}{\bibfnamefont{N.}~\bibnamefont{Gross}},
  \bibinfo{author}{\bibfnamefont{Z.}~\bibnamefont{Shotan}},
  \bibinfo{author}{\bibfnamefont{O.}~\bibnamefont{Machtey}},
  \bibinfo{author}{\bibfnamefont{S.}~\bibnamefont{Kokkelmans}},
  \bibnamefont{and}
  \bibinfo{author}{\bibfnamefont{L.}~\bibnamefont{Khaykovich}},
  \bibinfo{journal}{Comptes Rendus Phys.} \textbf{\bibinfo{volume}{12}},
  \bibinfo{pages}{4} (\bibinfo{year}{2011}).

\bibitem[{\citenamefont{Mies}({1968})}]{Mies1968}
\bibinfo{author}{\bibfnamefont{F.~H.} \bibnamefont{Mies}},
  \bibinfo{journal}{{Phys. Rev.}} \textbf{\bibinfo{volume}{{175}}},
  \bibinfo{pages}{{164}} (\bibinfo{year}{{1968}}).

\bibitem[{\citenamefont{Gross et~al.}(2009)\citenamefont{Gross, Shotan,
  Kokkelmans, and Khaykovich}}]{Gross2009}
\bibinfo{author}{\bibfnamefont{N.}~\bibnamefont{Gross}},
  \bibinfo{author}{\bibfnamefont{Z.}~\bibnamefont{Shotan}},
  \bibinfo{author}{\bibfnamefont{S.}~\bibnamefont{Kokkelmans}},
  \bibnamefont{and}
  \bibinfo{author}{\bibfnamefont{L.}~\bibnamefont{Khaykovich}},
  \bibinfo{journal}{Phys. Rev. Lett.} \textbf{\bibinfo{volume}{103}},
  \bibinfo{pages}{163202} (\bibinfo{year}{2009}).

\bibitem[{\citenamefont{Chin et~al.}(2004)\citenamefont{Chin,
  Vuleti\ifmmode~\acute{c}\else \'{c}\fi{}, Kerman, Chu, Tiesinga, Leo, and
  Williams}}]{Chin2004}
\bibinfo{author}{\bibfnamefont{C.}~\bibnamefont{Chin}},
  \bibinfo{author}{\bibfnamefont{V.}~\bibnamefont{Vuleti\ifmmode~\acute{c}\else
  \'{c}\fi{}}}, \bibinfo{author}{\bibfnamefont{A.~J.} \bibnamefont{Kerman}},
  \bibinfo{author}{\bibfnamefont{S.}~\bibnamefont{Chu}},
  \bibinfo{author}{\bibfnamefont{E.}~\bibnamefont{Tiesinga}},
  \bibinfo{author}{\bibfnamefont{P.~J.} \bibnamefont{Leo}}, \bibnamefont{and}
  \bibinfo{author}{\bibfnamefont{C.~J.} \bibnamefont{Williams}},
  \bibinfo{journal}{Phys. Rev. A} \textbf{\bibinfo{volume}{70}},
  \bibinfo{pages}{032701} (\bibinfo{year}{2004}).

\bibitem[{\citenamefont{Gross et~al.}(2010{\natexlab{b}})\citenamefont{Gross,
  Shotan, Kokkelmans, and Khaykovich}}]{Gross2010}
\bibinfo{author}{\bibfnamefont{N.}~\bibnamefont{Gross}},
  \bibinfo{author}{\bibfnamefont{Z.}~\bibnamefont{Shotan}},
  \bibinfo{author}{\bibfnamefont{S.}~\bibnamefont{Kokkelmans}},
  \bibnamefont{and}
  \bibinfo{author}{\bibfnamefont{L.}~\bibnamefont{Khaykovich}},
  \bibinfo{journal}{Phys. Rev. Lett.} \textbf{\bibinfo{volume}{105}},
  \bibinfo{pages}{103203} (\bibinfo{year}{2010}{\natexlab{b}}).

\bibitem[{\citenamefont{Yan et~al.}(1996)\citenamefont{Yan, Babb, Dalgarno, and
  Drake}}]{Yan1996}
\bibinfo{author}{\bibfnamefont{Z.-C.} \bibnamefont{Yan}},
  \bibinfo{author}{\bibfnamefont{J.~F.} \bibnamefont{Babb}},
  \bibinfo{author}{\bibfnamefont{A.}~\bibnamefont{Dalgarno}}, \bibnamefont{and}
  \bibinfo{author}{\bibfnamefont{G.~W.~F.} \bibnamefont{Drake}},
  \bibinfo{journal}{Phys. Rev. A} \textbf{\bibinfo{volume}{54}},
  \bibinfo{pages}{2824} (\bibinfo{year}{1996}).

\bibitem[{\citenamefont{Z\"urn et~al.}(2013)\citenamefont{Z\"urn, Lompe, Wenz,
  Jochim, Julienne, and Hutson}}]{Zurn2013}
\bibinfo{author}{\bibfnamefont{G.}~\bibnamefont{Z\"urn}},
  \bibinfo{author}{\bibfnamefont{T.}~\bibnamefont{Lompe}},
  \bibinfo{author}{\bibfnamefont{A.~N.} \bibnamefont{Wenz}},
  \bibinfo{author}{\bibfnamefont{S.}~\bibnamefont{Jochim}},
  \bibinfo{author}{\bibfnamefont{P.~S.} \bibnamefont{Julienne}},
  \bibnamefont{and} \bibinfo{author}{\bibfnamefont{J.~M.}
  \bibnamefont{Hutson}}, \bibinfo{journal}{Phys. Rev. Lett.}
  \textbf{\bibinfo{volume}{110}}, \bibinfo{pages}{135301}
  (\bibinfo{year}{2013}).

\bibitem[{Gau()}]{GaussNote}
\bibinfo{note}{Units of gauss rather than tesla, the accepted SI unit for the
  magnetic field, have been used in this paper to conform to the conventional
  usage in this field of physics.}

\bibitem[{not()}]{note1}
\bibinfo{note}{Following the usage of Ref.~\cite{Chin2010}, an $\ell$-wave
  resonance means an $\ell$-wave bound states is coupled to an $s$-wave open
  channel.}

\end{thebibliography}
\end{document}